# Policy and Planning for Large Infrastructure Projects:

# Problems, Causes, Cures

By

Bent Flyvbjerg

## 1. Introduction [*]

For a number of years my research group and I have explored different aspects of the planning of large infrastructure projects (Flyvbjerg, Bruzelius, and Rothengatter, 2003; Flyvbjerg, Holm, and Buhl, 2002, 2004, 2005; Flyvbjerg and Cowi, 2004; Flyvbjerg, 2005a, 2005b).[1] In this paper I would like to take stock of what we have learned from our research so far.

First I will argue that a major problem in the planning of large infrastructure projects is the high level of misinformation about costs and benefits that decision makers face in deciding whether to build, and the high risks such misinformation generates. Second I will explore the causes of misinformation and risk, mainly in the guise of optimism bias and strategic misrepresentation. Finally, I will present a number of measures aimed at improved planning and decision making, including changed incentive structures and better planning methods. Thus the paper is organized as a simple triptych consisting in problems, causes, and cures.

The emphasis will be on transportation infrastructure projects. I would like to mention at the outset, however, that comparative research shows that the problems, causes, and cures we identify for transportation apply to a wide range of other project types including power plants, dams, water projects, concert halls, museums, sports arenas, convention centers, IT systems, oil and gas extraction projects, aerospace projects, and weapons systems (Flyvbjerg, Bruzelius, and Rothengatter, 2003: 18-19; Flyvbjerg, Holm, and Buhl, 2002: 286; Flyvbjerg, 2005a; Altshuler and Luberoff, 2003).

## 2. Problems

Large infrastructure projects, and planning for such projects, generally have the following characteristics (Flyvbjerg and Cowi, 2004):

- Such projects are inherently risky due to long planning horizons and complex interfaces.
- Technology is often not standard.

---

[1] By "large infrastructure projects" I here mean the most expensive infrastructure projects that are built in the world today, typically at costs per project from around a hundred million to several billion dollars.

- Decision making and planning are often multi-actor processes with conflicting interests.
- Often the project scope or ambition level will change significantly over time.
- Statistical evidence shows that such unplanned events are often unaccounted for, leaving budget contingencies sorely inadequate.
- As a consequence, misinformation about costs, benefits, and risks is the norm.
- The result is cost overruns and/or benefit shortfalls with a majority of projects.

## 2.1 The Size of Cost Overruns and Benefit Shortfalls

For transportation infrastructure projects, Table 1 shows the inaccuracy of construction cost estimates measured as the size of cost overrun.[2] For rail, average cost overrun is 44.7 percent measured in constant prices. For bridges and tunnels, the equivalent figure is 33.8 percent, and for roads 20.4 percent. The difference in cost overrun between the three project types is statistically significant, indicating that each type should be treated separately (Flyvbjerg, Holm, and Buhl, 2002).

The large standard deviations shown in Table 1 are as interesting as the large average cost overruns. The size of the standard deviations demonstrate that uncertainty and risk regarding cost overruns are large, indeed.

The following key observations pertain to cost overruns in transportation infrastructure projects:

- 9 out of 10 projects have cost overrun.
- Overrun is found in the 20 nations and 5 continents covered by the study.

---

[2] The data are from the largest database of its kind. All costs are construction costs measured in constant prices. Cost overrun, also sometimes called "cost increase" or "cost escalation," is measured according to international convention as actual out-turn costs minus estimated costs in percent of estimated costs. Actual costs are defined as real, accounted construction costs determined at the time of project completion. Estimated costs are defined as budgeted, or forecasted, construction costs at the time of decision to build. For reasons explained in Flyvbjerg (2005b) the figures for cost overrun presented here must be considered conservative.-- Ideally financing costs, operating costs, and maintenance costs would also be included in a study of costs. It is difficult, however, to find valid, reliable, and comparable data on these types of costs across a large number of projects.

- Overrun is constant for the 70-year period covered by the study, estimates have not improved over time.

Table 1: Inaccuracy of transportation project cost estimates by type of project, in constant prices.

| Type of project | No. of cases (N) | Avg. cost overrun % | Standard deviation |
|---|---|---|---|
| Rail | 58 | 44.7 | 38.4 |
| Bridges and tunnels | 33 | 33.8 | 62.4 |
| Road | 167 | 20.4 | 29.9 |

Table 2 shows the inaccuracy of traffic forecasts for rail and road projects.[3] For rail, actual passenger traffic is 51.4 percent lower than estimated traffic on average. This is equivalent to an average overestimate in rail passenger forecasts of no less than 105.6 percent. The result is large benefit shortfalls for rail. For roads, actual vehicle traffic is on average 9.5 percent higher than forecasted traffic. We see that rail passenger forecasts are biased, whereas this is not the case for road traffic forecasts. The difference between rail and road is statistically significant at a high level. Again the standard deviations are large, indicating that forecasting errors vary widely across projects (Flyvbjerg, Holm, and Buhl, 2005; Flyvbjerg, 2005b).

The following observations hold for traffic demand forecasts:

- 84 percent of rail passenger forecasts are wrong by more than ±20 percent.
- 9 out of 10 rail projects have overestimated traffic.
- 50 percent of road traffic forecasts are wrong by more than ±20 percent.
- The number of roads with overestimated and underestimated traffic, respectively, is about the same.

---

[3] Following international convention, inaccuracy is measured as actual traffic minus estimated traffic in percent of estimated traffic. Rail traffic is measured as number of passengers; road traffic as number of vehicles. The base year for estimated traffic is the year of decision to build. The forecasting year is the first full year of operations (Flyvbjerg, 2005b).

- Inaccuracy in traffic forecasts are found in the 14 nations and 5 continents covered by the study.
- Inaccuracy is constant for the 30-year period covered by the study, forecasts have not improved over time.

We conclude that if techniques and skills for arriving at accurate cost and traffic forecasts have improved over time, these improvements have not resulted in an increase in the accuracy of forecasts.

Table 2: Inaccuracy in forecasts of rail passenger and road vehicle traffic.

| Type of project | No. of cases (N) | Avg. inaccuracy % | Standard deviation |
|-----------------|------------------|-------------------|--------------------|
| Rail | 25 | –51.4 | 28.1 |
| Road | 183 | 9.5 | 44.3 |

If we combine the data in tables 1 and 2, we see that for rail an average cost overrun of 44.7 percent combines with an average traffic shortfall of 51.4 percent.[4] For roads, an average cost overrun of 20.4 percent combines with a fifty-fifty chance that traffic is also wrong by more than 20 percent. As a consequence, cost benefit analyses and social and environmental impact assessments based on cost and traffic forecasts like those described above will typically be highly misleading.

## 2.2 Examples of Cost Overruns and Benefit Shortfalls

The list of examples of projects with cost overruns and/or benefit shortfalls is seemingly endless (Flyvbjerg, 2005a). Boston's Big Dig, otherwise known as the Central Artery/Tunnel Project, were 275 percent or US$11 billion over budget in constant dollars when it opened, and further overruns are accruing due to faulty construction. Actual costs for Denver's $5 billion International Airport were close to 200 percent higher than estimated costs. The overrun on the San Francisco-Oakland Bay Bridge retrofit was $2.5 billion, or more than 100 percent, even before construction started. The Copenhagen metro and many other urban

---

[4] For each of twelve urban rail projects, we have data for both cost overrun and traffic shortfall. For these projects average cost overrun is 40.3 percent; average traffic shortfall is 47.8 percent.

rail projects worldwide have had similar overruns. The Channel tunnel between the UK and France came in 80 percent over budget for construction and 140 percent over for financing. At the initial public offering, Eurotunnel, the private owner of the tunnel, lured investors by telling them that 10 percent "would be a reasonable allowance for the possible impact of unforeseen circumstances on construction costs."[5] Outside of transportation, the $4 billion cost overrun for the Pentagon spy satellite program and the over $5 billion overrun on the International Space Station are typical of defense and aerospace projects. Our studies show that large infrastructure and technology projects tend statistically to follow a pattern of cost underestimation and overrun. Many such projects end up financial disasters. Unfortunately, the consequences are not always only financial, as is illustrated by the NASA space shuttle. Here, the cooking of budgets to make this under-performing project look good on paper has been linked with shortchanged safety upgrades related to the deaths of seven astronauts aboard the Columbia shuttle in 2003 (Flyvbjerg 2004).

As for benefit shortfalls, consider Bangkok's US$2 billion Skytrain, a two-track elevated urban rail system designed to service some of the most densely populated areas from the air. The system is greatly oversized, with station platforms too long for its shortened trains. Many trains and cars sit in the garage, because there is no need for them. Terminals are too large, etc. The reason is that actual traffic turned out to be less than half that forecast (Flyvbjerg, Holm, and Buhl, 2005: 132). Every effort has been made to market and promote the train, but the project company has ended up in financial trouble. Even though urban rail is probably a good idea for a dense, congested, and air-polluted city like Bangkok, overinvesting in idle capacity is hardly the best way to use resources, especially in a developing nation in which capital for investment is particularly scarce. Such benefit shortfalls are common and have also haunted the Channel tunnel, the Los Angeles and Copenhagen metros, and Denver's International Airport.

Other projects with cost overruns and/or benefit shortfalls are, in North America: the F/A-22 fighter aircraft; FBI's Trilogy information system; Ontario's Pickering nuclear plant; subways in numerous cities, including Miami and Mexico City; convention centers in Houston, Los Angeles, and other cities; the Animas-La Plata water project; the Sacramento regional sewer system renewal; the Quebec Olympic stadium; Toronto's Sky Dome; the

---

[5] Quoted from "Under Water Over Budget," *The Economist*, October 7, 1989, 37–38.

Washington Public Power Supply System; and the Iraq reconstruction effort. In Europe: the Eurofighter military jet, the new British Library, the Millennium Dome, the Nimrod maritime patrol plane, the UK West Coast rail upgrade and the related Railtrack fiscal collapse, the Astute attack submarine, the Humber Bridge, the Tyne metro system, the Scottish parliament building, the French Paris Nord TGV, the Berlin-Hamburg maglev train, Hanover's Expo 2000, Athens' 2004 Olympics, Russia's Sakhalin-1 oil and gas project, Norway's Gardermo airport, the Øresund Bridge between Sweden and Denmark, and the Great Belt rail tunnel linking Scandinavia with continental Europe. In Australasia: Sydney's Olympic stadiums, Japan's Joetsu Shinkansen high-speed rail line, India's Sardar Sarovar dams, the Surat-Manor toll way project, Calcutta's metro, and Malaysia's Pergau dam. I end the list here only for reasons of space.

*2.3 Why Cost Overruns and Benefit Shortfalls Are a Problem*

Cost overruns and benefit shortfalls of the frequency and size described above are a problem for the following reasons:

- They lead to a Pareto-inefficient allocation of resources, i.e., waste.
- They lead to delays and further cost overruns and benefit shortfalls.
- They destabilize policy, planning, implementation, and operations of projects.
- The problem is getting bigger, because projects get bigger.

Let's consider each point in turn. First, an argument often heard in the planning of large infrastructure projects is that cost and benefit forecasts at the planning stage may be wrong, but if one assumes that forecasts are wrong by the same margin across projects, cost-benefit analysis would still identify the best projects for implementation. The ranking of projects would not be affected by the forecasting errors, according to this argument. However, the large standard deviations shown in tables 1 and 2 falsify this argument. The standard deviations show that cost and benefit estimates are not wrong by the same margin across projects; errors vary extensively and this will affect the ranking of projects. Thus we see that misinformation about costs and benefits at the planning stage is likely to lead to Pareto-inefficiency, because in terms of standard cost-benefit analysis decision makers are likely to implement inferior projects.

Second, cost overruns of the size described above typically lead to delays, because securing additional funding to cover overruns often takes time. In addition, projects may need to be re-negotiated or re-approved when overruns are large, as the data show they often are (Flyvbjerg, 2005a). In a separate study, we demonstrated that delays in transportation infrastructure implementation are very costly, increasing the percentage construction cost overrun measured in constant prices by 4.64 percentage points per year of delay incurred after the time of decision to build (Flyvbjerg, Holm, and Buhl, 2004). For a project of, say, US$8 billion--that is the size range of the Channel tunnel and about half the size of Boston's Big Dig--the expected average cost of delay would be approximately $370 million/year, or about $1 million/day.--Benefit shortfalls are an additional consequence of delays, because delays result in later opening dates and thus

extra months or years without revenues. Because many large infrastructure projects are loan-financed and have long construction periods, they are particularly sensitive to delays, as delays result in increased debt, increased interest payments, and longer payback periods.

Third, large cost overruns and benefit shortfalls tend to destabilize policy, planning, implementation, and operations. For example, after several overruns in the initial phase of the Sydney Opera House, the Parliament of New South Wales decided that every further 10 percent increase in the budget would need their approval. After this decision, the Opera House became a political football needing constant re-approval. Every overrun set off an increasingly menacing debate about the project, in Parliament and outside, with total cost overruns ending at 1,400 percent. The unrest drove the architect off the project, destroyed his career and oeuvre, and produced an Opera House unsuited for opera. Many other projects have experienced similar, if less spectacular, unrest, including the Channel Tunnel, Boston's Big Dig, and Copenhagen's metro.

Finally, as projects grow bigger, the problems with cost overruns and benefit shortfalls also grow bigger and more consequential (Flyvbjerg, Holm, and Buhl, 2004: 12). Some megaprojects are becoming so large in relation to national economies that cost overruns and benefit shortfalls from even a single project may destabilize the finances of a whole country or region. This occurred when the billion-dollar cost overrun on the 2004 Athens Olympics affected the credit rating of Greece and when benefit shortfalls hit Hong Kong's new $20 billion Chek Lap Kok airport after it opened in 1998. The desire to avoid

national fiscal distress has recently become an important driver in attempts at reforming the planning of large infrastructure projects, as we will see later.

*2.4 Policy Implications*

The policy implications of the results presented above are clear:

- Lawmakers, investors, and the public cannot trust information about costs, benefits, and risks of large infrastructure projects produced by promoters and planners of such projects.
- The current way of planning large infrastructure projects is ineffective in conventional economic terms, i.e., it leads to Pareto-inefficient investments.
- There is a strong need for reform in policy and planning for large infrastructure projects.

Before depicting what reform may look like in this expensive and consequential policy area, we will examine the causes of cost overruns and benefit shortfalls.

## 3. Causes

Three main types of explanation exist that claim to account for inaccuracy in forecasts of costs and benefits: technical, psychological, and political-economic explanations.

*3.1 Technical Explanations*

Technical explanations account for cost overruns and benefit shortfalls in terms of imperfect forecasting techniques, inadequate data, honest mistakes, inherent problems in predicting the future, lack of experience on the part of forecasters, etc. This is the most common type of explanation of inaccuracy in forecasts (Ascher, 1978; Flyvbjerg, Holm, and Buhl, 2002, 2005; Morris and Hough, 1987; Wachs, 1990). Technical error may be reduced or eliminated by developing better forecasting models, better data, and more experienced forecasters, according to this explanation.

*3.2 Psychological Explanations*

Psychological explanations account for cost overruns and benefit shortfalls in terms of what psychologists call the planning fallacy and optimism bias. Such explanations have been developed by Kahneman and Tversky (1979), Kahneman and Lovallo (1993), and Lovallo and Kahneman (2003). In the grip of the planning fallacy, planners and project promoters make decisions based on delusional optimism rather than on a rational weighting of gains, losses, and probabilities. They overestimate benefits and underestimate costs. They involuntarily spin scenarios of success and overlook the potential for mistakes and miscalculations. As a result, planners and promoters pursue initiatives that are unlikely to come in on budget or on time, or to ever deliver the expected returns.

Overoptimism can be traced to cognitive biases, that is, errors in the way the mind processes information. These biases are thought to be ubiquitous, but their effects can be tempered by simple reality checks, thus reducing the odds that people and organizations will rush blindly into unprofitable investments of money and time.

*3.3 Political-Economic Explanations*

Political-economic explanations see planners and promoters as deliberately and strategically overestimating benefits and underestimating costs when forecasting the outcomes of projects. They do this in order to increase the likelihood that it is their projects, and not the competition's, that gain approval and funding. Political-economic explanations have been set forth by Flyvbjerg, Holm, and Buhl (2002, 2005) and Wachs (1989,1990). According to such explanations planners and promoters purposely spin scenarios of success and gloss over the potential for failure. Again, this results in the pursuit of ventures that are unlikely to come in on budget or on time, or to deliver the promised benefits.

Strategic misrepresentation can be traced to political and organizational pressures, for instance competition for scarce funds or jockeying for position, and it is rational in this sense. If we now define a lie in the conventional fashion as making a statement intended to deceive others (Bok, 1979: 14; Cliffe et al., 2000: 3), we see that deliberate misrepresentation of costs and benefits is lying, and we arrive at one of the most basic explanations of lying that exists: Lying pays off, or at least political and economic agents believe it does. Where there is political pressure there is misrepresentation and lying, according to this explanation, but misrepresentation and lying can be moderated by measures of accountability.

*3.4 How Valid Are Explanations?*

How well does each of the three explanations of forecasting inaccuracy--technical, psychological, and political-economic --account for the data on cost overruns and benefit shortfalls presented earlier? This is the question to be answered in this section.

Technical explanations have, as mentioned, gained widespread credence among forecasters and planners (Ascher, 1978; Flyvbjerg, Holm, and Buhl, 2002, 2005). It turns out, however, that such credence could mainly be upheld because until now samples have been too small to allow tests by statistical methods. The data presented above, which come from the first large-sample study in the field, lead us to reject technical explanations of forecasting inaccuracy. Such explanations do not fit the data well. First, if misleading forecasts were truly caused by technical inadequacies, simple mistakes, and inherent problems with predicting the future, we would expect a less biased distribution of errors in forecasts around zero. In fact, we have found with high statistical significance that for four out of five distributions of forecasting errors, the distributions have a mean statistically different from zero. Only the data for inaccuracy in road traffic forecasts have a statistical distribution that seem to fit with explanations in terms of technical forecasting error. Second, if imperfect techniques, inadequate data, and lack of experience were main explanations of inaccuracies, we would expect an improvement in accuracy over time, since in a professional setting errors and their sources would be recognized and addressed through the refinement of data collection, forecasting methods, etc. Substantial resources have in fact been spent over several decades on improving data and methods. Still our data show that this has had no effect on the accuracy of forecasts. Technical factors, therefore, do not appear to explain the data. It is not so-called forecasting "errors" or their causes that need explaining. It is the fact that in a large majority of cases, costs are underestimated and benefits overestimated. We may agree with proponents of technical explanations that it is, for example, impossible to predict for the individual project exactly *which* geological, environmental, or safety problems will appear and make costs soar. But we maintain that it is possible to predict the risk, based on experience from other projects, *that* some such problems will haunt a project and how this will affect costs. We also maintain that such risk can and should be accounted for in forecasts of costs, but typically is not. For technical explanations to be valid, they would have to explain why forecasts are so consistent in ignoring cost and benefit risks over time, location, and project type.

Psychological explanations better fit the data. The existence of optimism bias in planners and promoters would result in actual costs being higher and actual benefits being lower than those forecasted. Consequently, the existence of optimism bias would be able to account, in whole or in part, for the peculiar bias found in most of our data. Interestingly, however, when you ask forecasters about causes for forecasting inaccuracies in actual forecasts, they do not mention optimism bias as a main cause of inaccuracy (Flyvbjerg, Holm, and Buhl, 2005: 138-140). This could of course be because optimism bias is unconscious and thus not reflected by forecasters. After all, there is a large body of experimental evidence for the existence of optimism bias (Buehler et al., 1994; Buehler, Griffin, and MacDonald, 1997; Newby-Clark et al. 2000). However, the experimental data are mainly from simple, non-professional settings. This is a problem for psychological explanations, because it remains an open question whether they are general and apply beyond such simple settings. Optimism bias would be an important and credible explanation of underestimated costs and overestimated benefits in infrastructure forecasting if estimates were produced by inexperienced forecasters, i.e., persons who were estimating costs and benefits for the first or second time and who were thus unknowing about the realities of infrastructure building and were not drawing on the knowledge and skills of more experienced colleagues. Such situations may exist and may explain individual cases of inaccuracy. But given the fact that in modern society it is a defining characteristic of professional expertise that it is constantly tested--through scientific analysis, critical assessment, and peer review--in order to root out bias and error, it seems unlikely that a whole profession of forecasting experts would continue to make the same mistakes decade after decade instead of learning from their actions. Learning would result in the reduction, if not elimination, of optimism bias, which would then result in estimates becoming more accurate over time. But our data clearly shows that this has not happened. The profession of forecasters would indeed have to be an optimistic--and non-professional--group to keep their optimism bias throughout the 70-year period our study covers for costs, and the 30-year period covered for patronage, and not learn that they were deceiving themselves and others by underestimating costs and overestimating benefits. This would account for the data, but is not a credible explanation. Therefore, on the basis of our data, we are led to reject optimism bias as a primary cause of cost underestimation and benefit overestimation.

Political-economic explanations and strategic misrepresentation account well for the systematic underestimation of costs and overestimation of benefits found in the data. A strategic estimate of costs would be low, resulting in cost overrun, whereas a strategic estimate of benefits would be high, resulting in benefit shortfalls. A key question for explanations in terms of strategic misrepresentation is whether estimates of costs and benefits are intentionally biased to serve the interests of promoters in getting projects started. This question raises the difficult issue of lying. Questions of lying are notoriously hard to answer, because a lie is making a statement intended to deceive others, and in order to establish whether lying has taken place, one must therefore know the intentions of actors. For legal, economic, moral, and other reasons, if promoters and planners have intentionally cooked estimates of costs and benefits to get a project started, they are unlikely to formally tell researchers or others that this is the case. Despite such problems, two studies exist that succeeded in getting forecasters to talk about strategic misrepresentation (Flyvbjerg and Cowi, 2004; Wachs 1990).

Flyvbjerg and Cowi (2004) interviewed public officials, planners, and consultants who had been involved in the development of large UK transportation infrastructure projects. A planner with a local transportation authority is typical of how respondents explained the basic mechanism of cost underestimation:

"You will often as a planner know the real costs. You know that the budget is too low but it is difficult to pass such a message to the counsellors [politicians] and the private actors. They know that high costs reduce the chances of national funding."

Experienced professionals like the interviewee know that outturn costs will be higher than estimated costs, but because of political pressure to secure funding for projects they hold back this knowledge, which is seen as detrimental to the objective of obtaining funding.

Similarly, an interviewee explained the basic mechanism of benefit overestimation:

"The system encourages people to focus on the benefits--because until now there has not been much focus on the quality of risk analysis and the robustness [of projects]. It is therefore important for project promoters to demonstrate all the benefits, also because the project promoters know that their project is up against other projects and competing for scarce resources."

Such a focus on benefits and disregard of risks and robustness may consist, for instance, in the discounting of spatial assimilation problems described by Priemus (forthcoming) elsewhere in this issue. Competition between projects and authorities creates political and organizational pressures that in turn create an incentive structure that makes it rational for project promoters to emphasize benefits and deemphasize costs and risks. A project that looks highly beneficial on paper is more likely to get funded than one that does not.

Specialized private consultancy companies are typically engaged to help develop project proposals. In general, the interviewees found that consultants showed high professional standard and integrity. But interviewees also found that consultants appeared to focus on justifying projects rather than critically scrutinizing them. A project manager explained:

> "Most decent consultants will write off obviously bad projects but there is a grey zone and I think many consultants in reality have an incentive to try to prolong the life of projects which means to get them through the business case. It is in line with their need to make a profit."

The consultants interviewed confirmed that appraisals often focused more on benefits than on costs. But they said this was at the request of clients and that for specific projects discussed "there was an incredible rush to see projects realized."

"One typical interviewee saw project approval as "passing the test" and precisely summed up the rules of the game like this:

> "It's all about passing the test [of project approval]. You are in, when you are in. It means that there is so much focus on showing the project at its best at this stage."

In sum, the UK study shows that strong interests and strong incentives exist at the project approval stage to present projects as favorably as possible, that is, with benefits emphasized and costs and risks deemphasized. Local authorities, local developers and land owners, local labor unions, local politicians, local officials, local MPs, and consultants all stand to benefit from a project that looks favorable on paper and they have little incentive to actively avoid bias in estimates of benefits, costs, and risks. National bodies, like certain parts of the Department for Transport and the Ministry of Finance who fund and oversee projects, may have an interest in more realistic appraisals, but so far they have had little success in achieving such realism, although the situation may be changing with the initiatives to curb bias set out in HM Treasury (2003) and Flyvbjerg and Cowi (2004).

The second study was carried out by Martin Wachs (1990, 1986). Wachs interviewed public officials, consultants, and planners who had been involved in transit planning cases in the US. He found that a pattern of highly misleading forecasts of costs and patronage could not be explained by technical errors, honest mistakes, or inadequate methods. In case after case, planners, engineers, and economists told Wachs that they had had to "revise" their forecasts many times because they failed to satisfy their superiors. The forecasts had to be cooked in order to produce numbers that were dramatic enough to gain federal support for the projects whether or not they could be fully justified on technical grounds. Wachs (1990: 144) recounts from his interviews:

> "One young planner, tearfully explained to me that an elected county supervisor had asked her to estimate the patronage of a possible extension of a light-rail (streetcar) line to the downtown Amtrak station. When she carefully estimated that the route might carry two to three thousand passengers per day, the supervisor directed her to redo her calculations in order to show that the route would carry twelve to fifteen thousand riders per day because he thought that number necessary to justify a federal grant for system construction. When she refused, he asked her superior to remove her from the project, and to get someone else to 'revise' her estimates."

In another typical case of cost underestimation and benefit overestimation, Wachs (1990: 144-145) gives the following account:

> "a planner admitted to me that he had reluctantly but repeatedly adjusted the patronage figures upward, and the cost figures downward to satisfy a local elected official who wanted to compete successfully for a federal grant. Ironically, and to the chagrin of that planner, when the project was later built, and the patronage proved lower and the costs higher than the published estimates, the same local politician was asked by the press to explain the outcome. The official's response was to say, 'It's not my fault; I had to rely on the forecasts made by our staff, and they seem to have made a big mistake here'."

Like in the UK study above, Wachs specifically interviewed consultants. He found, as one consultant put it, that "success in the consulting business requires the forecaster to adjust results to conform with the wishes of the client," and clients typically wish to see costs underestimated and benefits overestimated (1990: 151-152).

On the basis of his pioneering study, Wachs (1990: 145) concludes that forecasts of costs and benefits are presented to the public as instruments for deciding whether or not a project is to be undertaken, but they are actually instruments for getting public funds committed to a favored project. Wachs (1990: 146, 1986: 28) talks of "nearly universal abuse" of forecasting in this context, and he finds no indication that it takes place only in transit planning; it is common in all sectors of the economy where forecasting routinely plays an important role in policy debates, according to Wachs.

In conclusion, the UK and US studies arrive at results that are basically similar. Both studies account well for existing data on cost underestimation and benefit overestimation. Both studies falsify the notion that in situations with high political and organizational pressure the lowballing of costs and highballing of benefits is caused by non-intentional technical error or optimism bias. Both studies support the view that in such situations promoters and forecasters intentionally use the following formula in order to secure approval and funding for their projects:

Underestimated costs  + Overestimated benefits =  Project approval

Using this formula, and thus "showing the project at its best" as one interviewee said above, results in an inverted Darwinism, i.e., the "survival of the unfittest." It is not the best projects that get implemented, but the projects that look best on paper. And the projects that look best on paper are the projects with the largest cost underestimates and benefit overestimates, other things being equal. But these are the worst, or "unfittest," projects in the sense that they are the very projects that will encounter most problems during construction and operations in terms of the largest cost overruns, benefit shortfalls, and risks of non-viability. They have been designed like that.

## 4. Cures

As should be clear, the planning and implementation of large infrastructure projects stand in need of reform. Less deception and more honesty are needed in the estimation of costs and benefits if better projects are to be implemented. This is not to say that costs and benefits are or should be the only basis for deciding whether to build large infrastructure projects. Clearly, forms of rationality other than economic rationality are at work in most projects and are balanced in the broader frame of public decision making. But the costs and benefits of large infrastructure projects often run in the hundreds of millions of dollars, with risks correspondingly high. Without knowledge of such risks, decisions are likely to be flawed.

When contemplating what planners can do to help reform come about, we need to distinguish between two fundamentally different situations: (1) planners and promoters consider it important to get forecasts of costs, benefits, and risks right, and (2) planners and promoters do not consider it important to get forecasts right, because optimistic forecasts are seen as a necessary means to getting projects started. The first situation is the easier one to deal with and here better methodology will go a long way in improving planning and decision making. The second situation is more difficult, and more common as we saw above. Here changed incentives are essential in order to reward honesty and punish deception, where today's incentives often do the exact opposite.

Thus two main measures of reform are (1) better forecasting methods, and (2) improved incentive structures, with the latter being the more important.

### 4.1 Better Methods: Reference Class Forecasting

If planners genuinely consider it important to get forecasts right, we recommend they use a new forecasting method called "reference class forecasting" to reduce inaccuracy and bias. This method was originally developed to compensate for the type of cognitive bias in human forecasting that Princeton psychologist Daniel Kahneman found in his Nobel prize-winning work on bias in economic forecasting (Kahneman, 1994; Kahneman and Tversky, 1979). Reference class forecasting has proven more accurate than conventional forecasting. In April 2005, based on a study by Flyvbjerg, Holm, and Buhl (2005), the American Planning Association (2005) officially endorsed reference class forecasting:

"APA encourages planners to use reference class forecasting in addition to traditional methods as a way to improve accuracy. The reference class forecasting method is beneficial for non-routine projects such as stadiums, museums, exhibit centers, and other local one-off projects. Planners should never rely solely on civil engineering technology as a way to generate project forecasts."

For reasons of space, here we present only an outline of the method, based mainly on Lovallo and Kahneman (2003) and Flyvbjerg (2003). In a different context, we are currently developing what is, to our knowledge, the first instance of practical reference class forecasting in planning (Flyvbjerg and Cowi, 2004).

Reference class forecasting consists in taking a so-called "outside view" on the particular project being forecast. The outside view is established on the basis of information from a class of similar projects. The outside view does not try to forecast the specific uncertain events that will affect the particular project, but instead places the project in a statistical distribution of outcomes from this class of reference projects.

Reference class forecasting requires the following three steps for the individual project:

(1)     Identification of a relevant reference class of past projects. The class must be broad enough to be statistically meaningful but narrow enough to be truly comparable with the specific project.

(2)     Establishing a probability distribution for the selected reference class. This requires access to credible, empirical data for a sufficient number of projects within the reference class to make statistically meaningful conclusions.

(3)     Compare the specific project with the reference class distribution, in order to establish the most likely outcome for the specific project.

Daniel Kahneman relates the following story about curriculum planning to illustrate reference class forecasting in practice (Lovallo and Kahneman 2003: 61). We use this example, because similar examples do not exist as yet in the field of infrastructure planning. Some years ago, Kahneman was involved in a project to develop a curriculum for a new subject area for high schools in Israel. The project was carried out by a team of academics and teachers. In time, the team began to discuss how long the project would take to complete. Everyone on the team was asked to write on a slip of paper the number of months needed to finish and report the project. The estimates ranged from 18 to 30 months. One of the team members--a distinguished expert in curriculum development--was then posed a challenge by another team member to recall as many projects similar to theirs as possible and to think of these projects as they were in a stage comparable to their project. "How long did it take them at that point to reach completion?", the expert was asked. After a while he answered, with some discomfort, that not all the comparable teams he could think of ever did complete their task. About 40 percent of them eventually gave up. Of those remaining, the expert could not think of any that completed their task in less than seven years, nor of any that took more than ten. The expert was then asked if he had reason to believe that the present team was more skilled in curriculum development than the earlier ones had been. The expert said no, he did not see any relevant factor that distinguished this team favorably from the teams he had been thinking about. His impression was that the present team was slightly below average in terms of resources and potential. The wise decision at this point would probably have been for the team to break up, according to Kahneman. Instead, the members ignored the pessimistic information and proceeded with the project. They finally completed the project eight years later, and their efforts went largely wasted--the resulting curriculum was rarely used.

In this example, the curriculum expert made two forecasts for the same problem and arrived at very different answers. The first forecast was the inside view; the second was the outside view, or the reference class forecast. The inside view is the one that the expert and the other team members adopted. They made forecasts by focusing tightly on the case at hand, considering its objective, the resources they brought to it, and the obstacles to its completion. They constructed in their minds scenarios of their coming progress and extrapolated current trends into the future. The resulting forecasts, even the most conservative ones, were overly optimistic. The outside view is the one provoked by the question to the curriculum expert. It completely ignored the details of the project at hand, and it involved no attempt at forecasting the events that would influence the project's future course. Instead, it examined the experiences of a class of similar projects, laid out a rough distribution of outcomes for this reference class, and then positioned the current project in that distribution. The resulting forecast, as it turned out, was much more accurate.

Similarly--to take an example from our work with developing reference class forecasting for practical infrastructure planning--planners in a city preparing to build a new subway would, first, establish a reference class of comparable projects. This could be the relevant rail projects from the sample used for this article. Through analyses the planners would establish that the projects included in the reference class were indeed comparable. Second, if the planners were concerned, for example, with getting construction cost estimates right, they would then establish the distribution of outcomes for the reference class regarding the accuracy of construction cost forecasts.

**Figure 1**

**Inaccuracy of construction cost forecasts for rail projects in reference class.**

**Average cost increase is indicated for non-UK and UK projects, separately.**

**Constant prices.**

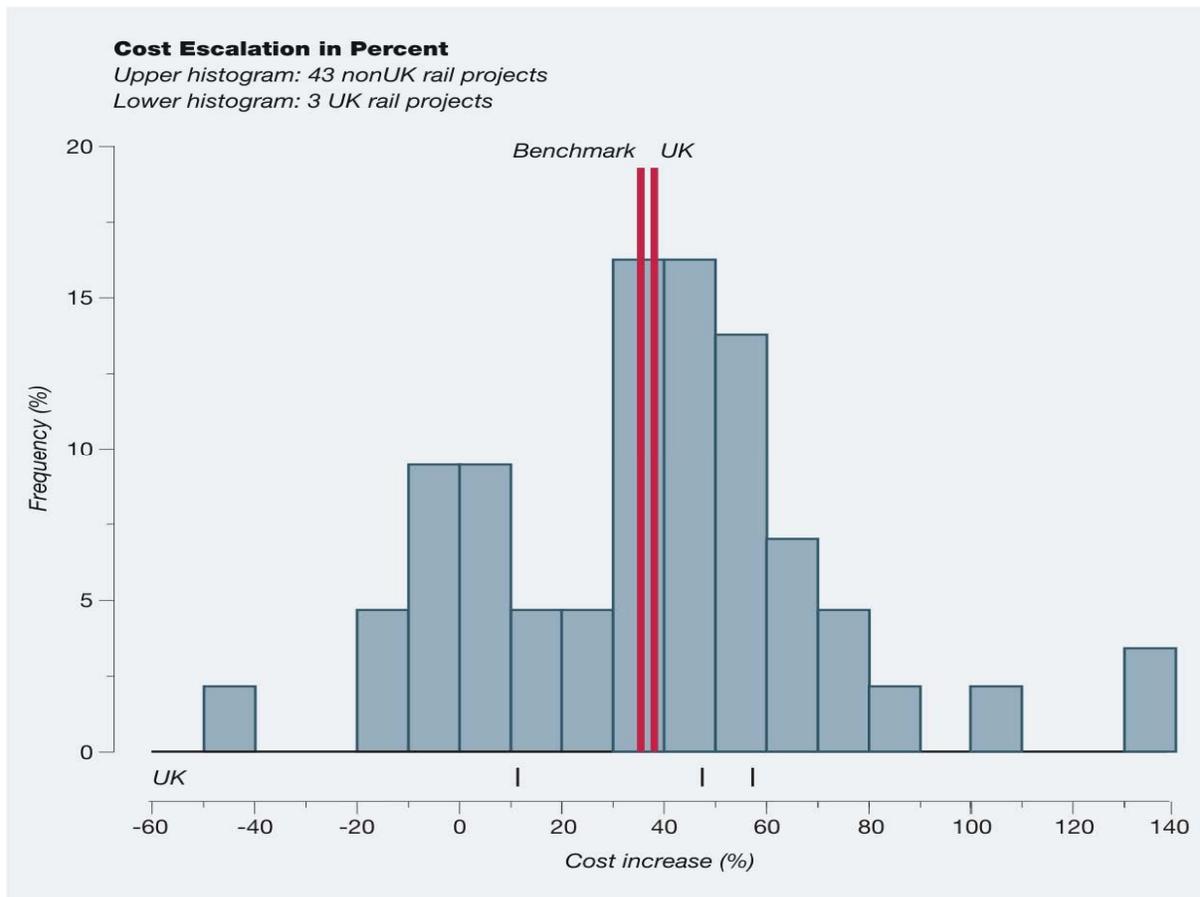

Figure 1 shows what this distribution looks like for a reference class relevant to building subways in the UK, developed by Flyvbjerg and Cowi (2004: 23) for the UK Department for Transport. Third, the planners would compare their subway project to the reference class distribution. This would make it clear to the planners that unless they have reason to believe they are substantially better forecasters and planners than their colleagues who did the forecasts and planning for projects in the reference class, they are likely to grossly underestimate construction costs. Finally, planners would then use this knowledge to adjust their forecasts for more realism.

**Figure 2:**

**Required adjustments to cost estimates for UK rail projects as function of the maximum acceptable level of risk for cost overrun.**

**Constant prices.**

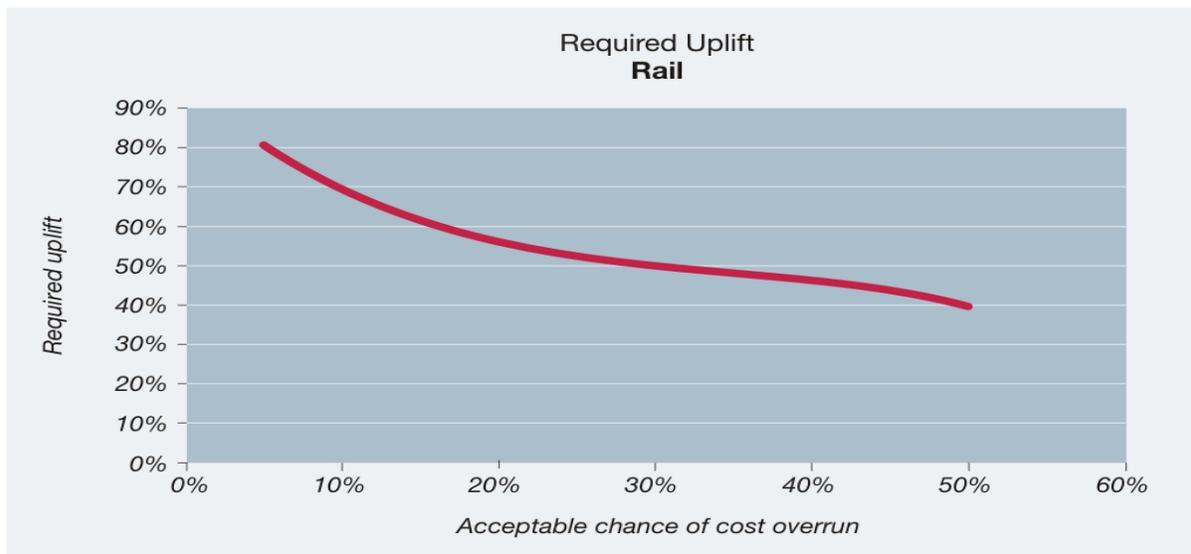

Figure 2 shows what such adjustments are for the UK situation. More specifically, Figure 2 shows that for a forecast of construction costs for a rail project, which has been planned in the manner that such projects are usually planned, i.e., like the projects in the reference class, this forecast would have to be adjusted upwards by 40 percent, if investors were willing to accept a risk of cost overrun of 50 percent. If investors were willing to accept a risk of overrun of only 10 percent, the uplift would have to be 68 percent. For a rail project initially estimated at, say £4 billion, the uplifts for the 50 and 10 percent levels of risk of cost overrun would be £1.6 billion and £2.7 billion, respectively.

The contrast between inside and outside views has been confirmed by systematic research (Gilovich, Griffin, and Kahneman, 2002). The research shows that when people are asked simple questions requiring them to take an outside view, their forecasts become significantly more accurate. However, most individuals and organizations are inclined to adopt the inside view in planning major initiatives. This is the conventional and intuitive approach. The traditional way to think about a complex project is to focus on the project itself and its details, to bring to bear what one knows about it, paying special attention to its unique or unusual features, trying to predict the events that will influence its future. The thought of going out and gathering simple statistics about related cases seldom enters a planner's mind. This is the case in general, according to Lovallo and Kahneman (2003: 61-62). And it is certainly the case for cost and benefit forecasting in large infrastructure projects. Despite the many forecasts we have reviewed, we have not come across a single genuine reference class forecast of costs and benefits.[6]

While understandable, planners' preference for the inside view over the outside view is unfortunate. When both forecasting methods are applied with equal skill, the outside view is much more likely to produce a realistic estimate. That is because it bypasses cognitive and political biases such as optimism bias and strategic misrepresentation and cuts directly to outcomes. In the outside view planners and forecasters are not required to make scenarios, imagine events, or gauge their own and others' levels of ability and control, so they cannot get all these things wrong. Surely the  outside view, being based on historical precedent, may fail to predict extreme outcomes, that is, those that lie outside all historical precedents. But for most projects, the outside view will produce more accurate results. In contrast, a focus on inside details is the road to inaccuracy.

The comparative advantage of the outside view is most pronounced for non-routine projects, understood as projects that planners and decision makers in a certain locale have never attempted before--like building an urban rail system in a city for the first time, or a new major bridge or tunnel where none existed before. It is in the planning of such new efforts that the biases toward optimism and strategic misrepresentation are likely to be largest. To be sure, choosing the right reference class of comparative past projects becomes more difficult when planners are forecasting initiatives for which precedents are not easily found, for instance the introduction of new and unfamiliar technologies. However, most large infrastructure projects are both non-routine locally and use well-known technologies. Such projects are, therefore, particularly likely to benefit from the outside view and reference class forecasting. The same holds for concert halls, museums, stadiums, exhibition centers, and other local one-off projects.

---

[6] The closest we have come to an outside view in large infrastructure forecasting is Gordon and Wilson's (1984) use of regression analysis on an international cross section of light-rail projects to forecast patronage in a number of light-rail schemes in North America.

*4.2 Improved Incentives: Public and Private Sector Accountability*

In the present section we consider the situation where planners and other influential actors do not find it important to get forecasts right and where planners, therefore, do not help to clarify and mitigate risk but, instead, generate and exacerbate it. Here planners are part of the problem, not the solution. This situation may need some explication, because it possibly sounds to many like an unlikely state of affairs. After all, it may be agreed that planners ought to be interested in being accurate and unbiased in forecasting. It is even stated as an explicit requirement in the AICP Code of Ethics and Professional Conduct that "A planner must strive to provide full, clear and accurate information on planning issues to citizens and governmental decision-makers" (American Planning Association, 1991: A.3). The British RTPI has laid down similar obligations for its members (Royal Town Planning Institute, 2001).

However, the literature is replete with things planners and planning "must" strive to do, but which they don't. Planning must be open and communicative, but often it is closed. Planning must be participatory and democratic, but often it is an instrument of domination and control. Planning must be about rationality, but often it is about power (Flyvbjerg, 1998; Watson, 2003). This is the "dark side" of planning and planners identified by Flyvbjerg (1996) and Yiftachel (1998), which is remarkably underexplored by planning researchers and theorists.

Forecasting, too, has its dark side. It is here that "planners lie with numbers," as Wachs (1989) has aptly put it. Planners on the dark side are busy not with getting forecasts right and following the AICP Code of Ethics but with getting projects funded and built. And accurate forecasts are often not an effective means for achieving this objective. Indeed, accurate forecasts may be counterproductive, whereas biased forecasts may be effective in competing for funds and securing the go-ahead for construction. "The most effective planner," says Wachs (1989: 477), "is sometimes the one who can cloak advocacy in the guise of scientific or technical rationality." Such advocacy would stand in direct opposition to AICP's ruling that "the planner's primary obligation [is] to the public interest" (American Planning Association, 1991: B.2). Nevertheless, seemingly rational forecasts that underestimate costs and overestimate benefits have long been an established formula for project approval as we saw above. Forecasting is here mainly another kind of rent-seeking behavior, resulting in a make-believe world of misrepresentation which makes it extremely difficult to decide which projects deserve undertaking and which do not. The consequence is, as even one of the industry's own organs, the Oxford-based Major Projects Association, acknowledges, that too many projects proceed that should not. We would like to add that many projects don't proceed that probably should, had they not lost out to projects with "better" misrepresentation (Flyvbjerg, Holm, and Buhl, 2002).

In this situation, the question is not so much what planners can do to reduce inaccuracy and risk in forecasting, but what others can do to impose on planners the checks and balances that would give planners the incentive to stop producing biased forecasts and begin to work according to their Code of Ethics. The challenge is to change the power relations that govern forecasting and project development. Better forecasting techniques and appeals to ethics won't do here; institutional change with a focus on transparency and accountability is necessary.

As argued in Flyvbjerg, Bruzelius, and Rothengatter (2003), two basic types of accountability define liberal democracies: (1) public sector accountability through transparency and public control, and (2) private sector accountability via competition and market control. Both types of accountability may be effective tools to curb planners' misrepresentation in forecasting and to promote a culture which acknowledges and deals effectively with risk. In order to achieve accountability through *transparency and public control*, the following would be required as practices embedded in the relevant institutions (the full argument for the measures may be found in Flyvbjerg, Bruzelius, and Rothengatter, 2003, chapters 9-11):

•       National-level government should not offer discretionary grants to local infrastructure agencies for the sole purpose of building a specific type of infrastructure. Such grants create perverse incentives. Instead, national government should simply offer "infrastructure grants" or "transportation grants" to local governments, and let local political officials spend the funds however they choose to, but make sure that every dollar they spend on one type of infrastructure reduces their ability to fund another.

- Forecasts should be made subject to independent peer review. Where large amounts of taxpayers' money are at stake, such review may be carried out by national or state accounting and auditing offices, like the General Accounting Office in the US or the National Audit Office in the UK, who have the independence and expertise to produce such reviews. Other types of independent review bodies may be established, for instance within national departments of finance or with relevant professional bodies.

- Forecasts should be benchmarked against comparable forecasts, for instance using reference class forecasting as described in the previous section.

- Forecasts, peer reviews, and benchmarkings should be made available to the public as they are produced, including all relevant documentation.

- Public hearings, citizen juries, and the like should be organized to allow stakeholders and civil society to voice criticism and support of forecasts. Knowledge generated in this way should be integrated in planning and decision making.

- Scientific and professional conferences should be organized where forecasters would present and defend their forecasts in the face of colleagues' scrutiny and criticism.

- Projects with inflated benefit-cost ratios should be reconsidered and stopped if recalculated costs and benefits do not warrant implementation. Projects with realistic estimates of benefits and costs should be rewarded.

- Professional and occasionally even criminal penalties should be enforced for planners and forecasters who consistently and foreseeably produce deceptive forecasts. An example of a professional penalty would be the exclusion from one's professional organization if one violates its code of ethics. An example of a criminal penalty would be punishment as the result of prosecution before a court or similar legal set-up, for instance where deceptive forecasts have led to substantial mismanagement of public funds (Garett and Wachs, 1996). Malpractice in planning should be taken as seriously as it is in other professions. Failing to do this amounts to not taking the profession of planning seriously.

In order to achieve accountability in forecasting via *competition and market control*, the following would be required, again as practices that are both embedded in and enforced by the relevant institutions:

- The decision to go ahead with a project should, where at all possible, be made contingent on the willingness of private financiers to participate without a sovereign guarantee for at least one third of the total capital needs.[7] This should be required whether projects pass the market test or not, that is, whether projects are subsidized or not or provided for social justice reasons or not. Private lenders, shareholders, and stock market analysts would produce their own forecasts or would critically monitor existing ones. If they were wrong about the forecasts, they and their organizations would be hurt. The result would be more realistic forecasts and reduced risk.
- Full public financing or full financing with a sovereign guarantee should be avoided.
- Forecasters and their organizations must share financial responsibility for covering cost overruns and benefit shortfalls resulting from misrepresentation and bias in forecasting.
- The participation of risk capital should not mean that government gives up or reduces control of the project. On the contrary, it means that government can more effectively play the role it should be playing, namely as the ordinary citizen's guarantor for ensuring concerns about safety, environment, risk, and a proper use of public funds.

Whether projects are public, private, or public-private, they should be vested in one and only one project organization with a strong governance framework. The project organization may be a company or not, public or private, or a mixture. What is important is that this organization enforces accountability vis-à-vis contractors, operators, etc., and that, in turn, the directors of the organization are held accountable for any cost overruns, benefits shortfall, faulty designs, unmitigated risks, etc. that may occur during project planning, implementation, and operations.

---

[7] The lower limit of a one-third share of private risk capital for such capital to effectively influence accountability is based on practical experience. See more in Flyvbjerg, Bruzelius, and Rothengatter (2003: 120-123).

If the institutions with responsibility for developing and building major infrastructure projects would effectively implement, embed, and enforce such measures of accountability, then the misrepresentation in cost, benefit, and risk estimates, which is widespread today, may be mitigated. If this is not done, misrepresentation is likely to continue, and the allocation of funds for infrastructure is likely to continue to be wasteful and undemocratic.

## 5. Toward Better Practice

Fortunately, after decades of widespread mismanagement of the planning and design of large infrastructure projects, signs of improvement have recently appeared. The conventional consensus that deception is an acceptable way of getting projects started is under attack, as will be apparent from the examples below. This is in part because democratic governance is generally getting stronger around the world. The Enron scandal and its successors have triggered a war on corporate deception that is spilling over into government with the same objective: to curb financial waste and promote good governance. Although progress is slow, good governance is gaining a foothold even in large infrastructure project development. The conventional consensus is also under attack for the practical reason mentioned earlier that the largest projects are now so big in relation to national economies that cost overruns, benefit shortfalls, and risks from even a single project may destabilize the finances of a whole country or region, as happened in Greece and Hong Kong. Lawmakers and governments begin to see that national fiscal distress is too high a price to pay for the conventional way of planning and designing large projects. The main drive for reform comes from outside the agencies and industries conventionally involved in infrastructure development, which increases the likelihood of success.

In 2003 the Treasury of the United Kingdom required, for the first time, that all ministries develop and implement procedures for large public projects that will curb what it calls--with true British civility--"optimism bias." Funding will be unavailable for projects that do not take into account this bias, and methods have been developed for how to do this (Mott MacDonald, 2002; HM Treasury, 2003; Flyvbjerg and Cowi, 2004). In the Netherlands in 2004, the Parliamentary Committee on Infrastructure Projects for the first time conducted extensive public hearings to identify measures that will limit the media (misinformation about large infrastructure projects given to the Parliament, public, and Tijdelijke Commissie Infrastructuurprojecten, 2004). In Boston, the government sued to recoup funds from contractor overcharges for the Big Dig related to cost overruns. More governments and parliaments are likely to follow the lead of the UK, the Netherlands, and Boston in coming years. It's too early to tell whether the measures they implement will ultimately be effective. It seems unlikely, however, that the forces that have triggered the measures will be reversed, and it is those forces that reform-minded groups need to support and work with in order to curb deception and waste. This is the "tension-point" where convention meets reform, power-balances change, and new things may happen.

The key weapons in the war on deception and waste are accountability and critical questioning. The professional expertise of planners, engineers, architects, economists, and administrators is certainly indispensable to constructing the infrastructures that make society work. Our studies show, however, that the claims about costs, benefits, and risks made by these groups usually cannot be trusted and should be carefully examined by independent specialists and organizations. The same holds for claims made by project-promoting politicians and officials. Institutional checks and balances--including financial, professional, or even criminal penalties for consistent and unjustifiable biases in claims and estimates of costs, benefits, and risks--should be developed and employed. The key principle is that the cost of making a wrong forecast should fall on those making the forecast, a principle often violated today.

Many of the public-private partnerships currently emerging in large infrastructure projects contain more and better checks and balances than previous institutional setups, as has been demonstrated by the UK National Audit Office (2003). This is a step in the right direction but should be no cause for repose. All available measures for improvement must be employed. The conventional mode of planning and designing infrastructure has long historical roots and is deeply ingrained in professional and institutional practices. It would be naive to think it is easily toppled. Given the stakes involved--saving taxpayers from billions of dollars of waste, protecting citizens' trust in democracy and the rule of law, avoiding the destruction of spatial and environmental assets--this shouldn't deter us from trying.

# References

Altshuler, A. and D. Luberoff, 2003, *Mega-Projects: The Changing Politics of Urban Public Investment* (Washington, DC: Brookings Institution).

American Planning Association, 1991, AICP Code of Ethics and Professional Conduct. Adopted October 1978, as amended October 1991, http://www.planning.org

American Planning Association, 2005, "JAPA Article Calls on Planners to Help End Inaccuracies in Public Project Revenue Forecasting" http://www.planning.org/newsreleases/2005/ftp040705.htm, April 7

Ascher, W.,1979, *Forecasting: An appraisal for policy-makers and planners* (Baltimore: The Johns Hopkins University Press)

Bok, S., 1979, *Lying: moral choice in public and private life* (New York: Vintage)

Buehler, R., Griffin, D., and MacDonald, H., 1997, The role of motivated reasoning in optimistic time predictions. *Personality and Social Psychology Bulletin,* 23, 3, pp 238-247

Buehler, R., Griffin, D., and Ross, M., 1994, Exploring the "planning fallacy": Why people underestimate their task completion times. *Journal of Personality and Social Psychology,* 67, pp 366-381

Cliffe, L., Ramsey, M., & Bartlett, D., 2000, *The politics of lying: Implications for democracy* (London: Macmillan)

Flyvbjerg, B., 1996, The dark side of planning: Rationality and *Realrationalität*. In S. Mandelbaum, L. Mazza, and R. Burchell (Eds.), *Explorations in Planning Theory* (New Brunswick, NJ: Center for Urban Policy Research Press), pp 383-394

Flyvbjerg, B., 1998, *Rationality and power: Democracy in practice* (Chicago: University of Chicago Press)

Flyvbjerg, B., 2003, Delusions of Success: Comment on Dan Lovallo and Daniel Kahneman. *Harvard Business Review*, December 2003, pp 121-122

Flyvbjerg, Bent, 2005a, "Design by Deception: The Politics of Megaproject Approval." *Harvard Design Magazine*, no. 22, Spring/Summer, pp 50-59

Flyvbjerg, Bent, 2005b, "Measuring Inaccuracy in Travel Demand Forecasting: Methodological Considerations Regarding Ramp Up and Sampling." *Transportation Research A*, vol. 39, no. 6, pp 522-530

Flyvbjerg, Bent, Nils Bruzelius, and Werner Rothengatter, 2003, *Megaprojects and Risk: An Anatomy of Ambition* (Cambridge University Press)

Flyvbjerg, Bent, 2004, "Megaprojects and Risk: A Conversation With Bent Flyvbjerg." Interview conducted by Renia Ehrenfeucht. *Critical Planning*, vol. 11, 2004, pp. 51-63.

Flyvbjerg, Bent and Cowi, 2004, *Procedures for Dealing with Optimism Bias in Transport Planning: Guidance Document* (London: UK Department for Transport)

Flyvbjerg, Bent, Mette K. Skamris Holm, and Søren L. Buhl, 2002, "Underestimating Costs in Public Works Projects: Error or Lie?" *Journal of the American Planning Association*, vol. 68, no. 3, Summer, pp 279-295

Flyvbjerg, Bent, Mette K. Skamris Holm, and Søren L. Buhl, 2004, "What Causes Cost Overrun in Transport Infrastructure Projects?" *Transport Reviews*, vol. 24, no. 1, pp 3-18

Flyvbjerg, Bent, Mette K. Skamris Holm, and Søren L. Buhl, 2005, "How (In)accurate

Are Demand Forecasts in Public Works Projects? The Case of Transportation." *Journal of the American Planning Association*, vol. 71, no. 2, Spring, pp 131-146

Garett, M. and Wachs, M., 1996, *Transportation Planning on Trial: The Clean Air Act and Travel Forecastin* (Thousand Oaks, CA: Sage)

Gilovich, T., Griffin, D., and Kahneman, D., 2002, Eds., *Heuristics and Biases: The Psychology of Intuitive Judgment* (Cambridge University Press, 2002)

Gordon, P. and Wilson, R., 1984, "The Determinants of Light-Rail Transit Demand: An International Cross-Sectional Comparison." *Transportation Research A*, 18A(2), pp 135-140

HM Treasury, 2003, *The Green Book: Appraisal and Evaluation in Central Government, Treasury Guidance* (London: TSO)

Kahneman, D., 1994, New challenges to the rationality assumption. *Journal of Institutional and Theoretical Economics*, 150, pp 18-36

Kahneman, D. and Lovallo, D., 1993, Timid choices and bold forecasts: A cognitive perspective on risk taking. *Management Science*, 39, pp 17-31

Kahneman, D. and Tversky, A., 1979, "Prospect theory: An analysis of decisions under risk." *Econometrica*, 47, pp 313-327

Lovallo, Dan and Daniel Kahneman, 2003, "Delusions of Success: How Optimism Undermines Executives' Decisions," *Harvard Business Review*, July, pp 56-63

Morris, Peter W. G. and George H. Hough, 1987, *The Anatomy of Major Projects: A Study of the Reality of Project Management* (New York: John Wiley and Sons)

Mott MacDonald, 2002, *Review of Large Public Procurement in the UK*, study for HM

Treasury (London: HM Treasury)

National Audit Office, 2003, *PFI: Construction Performance*, report by the Comptroller and Auditor General, HC 371 Session 2002–2003: February 5, 2003 (London: National Audit Office)

Newby-Clark, I. R., McGregor, I., and Zanna, M. P., 2002, Thinking and caring about cognitive inconsistency: when and for whom does attitudinal ambivalence feel uncomfortable? *Journal of Personality and Social Psychology,* 82*,* pp 157-166

Priemus, Hugo, forthcoming, "Design of Large Infrastructure Projects: Disregarded Alternatives and Issues of Spatial Planning." *Environment and Planning B.*

Royal Town Planning Institute, 2001, Code of professional conduct. As last amended by the Council on 17 January 2001, http://www.rtpi.org.uk

Tijdelijke Commissie Infrastructuurprojecten, *Grote Projecten Uitvergroot: Een Infrastructuur voor Besluitvorming* (The Hague: Tweede Kamer der Staten-Generaal, 2004).

Wachs, M., 1986, Technique vs. advocacy in forecasting: A study of rail rapid transit. *Urban Resources*, 4(1), pp 23-30

Wachs, M., 1989, When Planners Lie with Numbers. *Journal of the American Planning Association*, 55(4), pp 476-479

Wachs, M., 1990, Ethics and advocacy in forecasting for public policy. *Business and Professional Ethics Journal*, 9(1-2), pp 141-157

Watson, V., 2003, Conflicting Rationalities: Implications for Planning Theory and Ethics. *Planning Theory and Practice*, 4(4), pp 395-408

Yiftachel, O., 1998, Planning and Social Control: Exploring the Dark Side. *Journal of Planning Literature*, 12(4), pp 395-406